\documentclass[journal]{IEEEtran}

\usepackage{amsmath}
\usepackage{amssymb}
\usepackage{psfrag}
\renewcommand{\vec}[1]{\ensuremath{\boldsymbol{#1}}}
\DeclareMathOperator{\Tr}{Tr} 

\usepackage{graphicx}
\usepackage{epstopdf}
\usepackage[tight,footnotesize]{subfigure}
\usepackage{cite}
\usepackage{todonotes}
\begin{document}
\title{Massive MIMO in Real Propagation Environments: Do All Antennas Contribute Equally?}
\author{\IEEEauthorblockN{Xiang Gao\IEEEauthorrefmark{1}, Ove Edfors\IEEEauthorrefmark{1}, Fredrik Tufvesson\IEEEauthorrefmark{1}, Erik G. Larsson\IEEEauthorrefmark{2}\\}
\IEEEauthorblockA{\IEEEauthorrefmark{1}Department of Electrical Information and Technology, Lund University, Sweden\\\IEEEauthorrefmark{2}Department of Electrical Engineering (ISY), Link{\"o}ping University, Sweden}}

\maketitle
\begin{abstract}
Massive MIMO can greatly increase both spectral and transmit-energy efficiency. 
This is achieved by allowing the number of antennas and RF chains to grow very large. 
However, the challenges include high system complexity and hardware energy consumption. 
Here we investigate the possibilities to reduce the required number of RF chains, 
by performing antenna selection. 
While this approach is not a very effective strategy for theoretical independent Rayleigh fading channels, 
a substantial reduction in the number of RF chains can be achieved for real massive MIMO channels, 
without significant performance loss. 
%This is due to large-scale fading over the array and differences in antenna patterns, 
%which make signals from certain antennas more valuable than others. 
We evaluate antenna selection performance on measured channels at 2.6~GHz, 
using a linear and a cylindrical array, both having 128 elements. 
Sum-rate maximization is used as the criterion for antenna selection.
A selection scheme based on convex optimization is nearly optimal and used as a benchmark.
The achieved sum-rate is compared with that of a very simple scheme
that selects the antennas with the highest received power. 
The power-based scheme gives performance close to the convex optimization scheme,
for the measured channels.
This observation indicates a potential for significant reductions of massive MIMO implementation complexity, 
by reducing the number of RF chains and performing antenna selection using simple algorithms.
\end{abstract}

\section{Introduction}\label{sec:intro}

Massive MIMO \cite{Marzetta_unlimited_bs_ant, scale_up_mimo, Massive_MIMO_Larsson2013, Lu2014_Overview_MA, Marzetta2015_intro_MA} is an emerging technology in wireless access.
By using a large number (tens to hundreds) of antennas at the base station,
and serving many users in the same time-frequency resource,
massive MIMO can improve the spectral and transmit-energy efficiency of conventional MIMO by orders of magnitude 
\cite{EE_Ngo, Huh2012_notsolarge, Guthy2013, Bjornson2013_MA_smallCell},
and simple signal processing schemes are expected to achieve near-optimal performance 
\cite{xiang_vtc, Yang2013_MF_ZF, Hoydis2013_HowMany}.
The basic premise of massive MIMO is that,
as confirmed by several experiments \cite{Xiang_Asilomar, Gao_MassiveMIMO_Real_Channel_2014, Hoydis_meas_2012, Jose_ICC_2014}, the propagation channel has a large number of spatial degrees of freedom. 
Massive MIMO is currently considered a leading 5G technology candidate \cite{5G_2014, Wang2014, Boccardi2014, Osseiran2014, Jungnickel2014}.
Real-time massive MIMO testbeds are being implemented and demonstrations are also reported 
\cite{Vieira2014_LuMami, Shepard2012_agos, Shepard2013_agosV2, Suzuki2012}. 
However, with a large number of antennas and associated transceiver chains, 
the challenges of massive MIMO include high system complexity and hardware power consumption 
\cite{mami_ee, Mohammed2014, Bjornson2015_ee, Liu2015_EE}.

This paper investigates whether all antennas in a massive MIMO system contribute equally to the overall performance or not.  
Experimental data from measurement campaigns at the 2.6~GHz band are used to demonstrate that in many cases, the antennas do \emph{not} contribute equally. This observation paves the way for antenna selection algorithms
and for hardware architectures where the number of activated radio-frequency (RF) transceiver chains is less than the
actual number of antennas. Antenna selection algorithms for such architectures are then proposed and their performance is analyzed. 
The practical impact of the proposed techniques is that the overall energy efficiency of massive MIMO systems can be substantially improved, and the hardware complexity can be reduced.

The rest of the paper is organized as follows. 
In Sec.~\ref{sec:background} we discuss the general background, 
and introduce the antenna selection concept.
Sec.~\ref{sec:approach} outlines the approach we have chosen for the study.
In Sec.~\ref{sec:system_model} we describe the system model and present two antenna selection schemes.
In Sec.~\ref{sec:measurements} we describe the channel measurement setup used to obtain the experimental results.
Then in Sec.~\ref{sec:results} we present performance results with antenna selection, 
and discuss the effectiveness of the proposed algorithms and how many transceiver chains that are needed under different operating conditions.
Conclusions are given in Sec.~\ref{sec:conclusions}.

\section{Background}\label{sec:background}

In ``ideal'' independent and identically distributed (i.i.d.) Rayleigh fading  channels,
all the antennas can be expected to contribute equally to the system performance.
To see why, consider a multi-user MIMO-OFDM system with $L$ subcarriers and suppose the base station has an array with $M$ antennas that serves $K$ users. 
Denote the $M\!\times\!1$ channel vector for a given user $k$ and a given subcarrier $\ell$ by $\vec{g}_k(\ell)$.
In i.i.d. Rayleigh fading channels, all antennas are equally good in the sense that 
\begin{equation}\label{eq:average_out}
	\frac{1}{K}\!\sum_{k=1}^{K}{\!\frac{1}{L}\!\sum_{\ell=1}^L{\left|\vec{g}_{k,m}(\ell)\right|^2}}\!\approx\!\mbox{constant\qquad for all $m$},
\end{equation}
where the constant is independent of the antenna index $m$.
This means that provided the bandwidth is relatively large and the number of users is large, no antenna outperforms the others.
%\todo[inline]{Ove: I'm not 100\% satisfied with the term "spatial diversity". Here it is quite understandable when combined with "over base station antennas", but in other places in the text I think it may be misinterpreted. Don't have a good solution though. Let's discuss.\\
%Xiang: I changed to ``the effectiveness of the diversity'' according to Fredrik's comments.}
 
\begin{figure}
  \centering
  \includegraphics[width=0.48\textwidth]{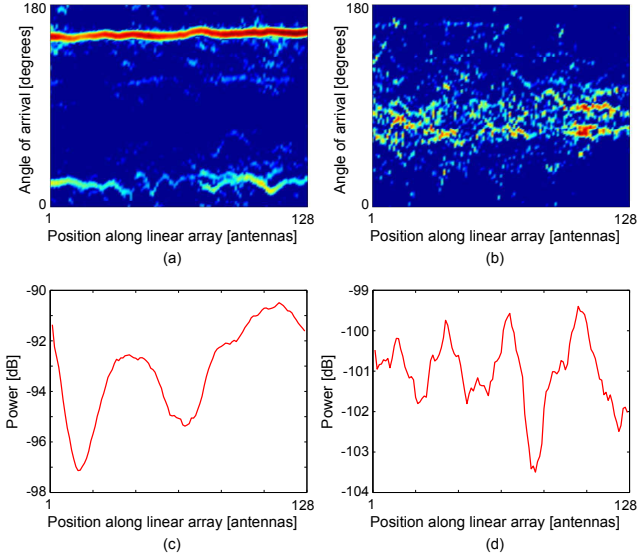} %- for double-column
  \caption{Angular power spectrum and power variation over a 7.4~m linear array, 
  in the measured channels as reported in \cite{Xiang_Asilomar_Channel_Model} and \cite{xiang_cost_2012}.
  The four plots show:
  (a) angular power spectrum in a LOS scenario, (b) angular power spectrum in a NLOS scenario,
  (c) average power variation in the LOS scenario, (d) average power variation in the NLOS scenario.}
  \label{fig:APS_both}
\end{figure}

In real propagation channels, however, the situation is different. 
Here, all the antennas contribute, but some antennas contribute more than others.
In the study based on measured channels at the 2.6~GHz band, 
using a linear array of omni-directional antennas and a cylindrical array of patch antennas, both having 128 elements,
we have observed that over the measured 50~MHz bandwidth the average power variations across the two arrays can be significant
\cite{Xiang_Asilomar_Channel_Model, Xiang_Asilomar_Interation, xiang_cost_2012}.
As an example, the angular power spectrum (APS) and power variation over the 7.4~m linear array are shown in Fig.~\ref{fig:APS_both},
for a line-of-sight (LOS) scenario and a non-line-of-sight (NLOS) scenario, respectively.
Unlike in conventional MIMO (thought of here as up to 8 antennas as in LTE \cite{3GPP_LTE_A}), 
the characteristics of the propagation channel across the linear array vary significantly.
In Fig.~\ref{fig:APS_both}(a) and Fig.~\ref{fig:APS_both}(b), 
some scatterers are not visible over the whole array, and for scatterers that are visible over the whole array, 
the power contributions vary considerably.
Consequently, in Fig.~\ref{fig:APS_both}(c) and Fig.~\ref{fig:APS_both}(d), we observe large power variations over the array,
about 7~dB in the LOS scenario and 4~dB in the NLOS scenario.
Thus, large-scale fading is experienced over the array.
The compact cylindrical array, which is smaller in size, 
experiences a similar effect of power variation over the array.
This is, however, due to its circular structure and patch antenna arrangement, 
rather than large-scale fading.
In contrast to i.i.d. Rayleigh fading channels,
in real massive MIMO channels the large power variation makes some antennas more ``useful'' than others,
and this power variation persists when averaging over frequency provided that the system is moderately wideband.

Since all antennas are not equally good in real propagation channels,
it is possible to reduce the number of active antennas and transceivers,
by selecting those that contribute the most and discarding the rest.
Such antenna selection could simplify the design of a massive MIMO base station and lead to energy and cost savings. 
One possible implementation is to deploy a large number of antennas but fewer RF transceivers,
exploiting the fact that antennas are relatively cheap while RF chains are expensive and energy consuming.
In this case, we need an RF switch,
which can be highly complex to implement and introduces losses in signal quality,  
especially when there are many antennas and transceivers.
Another implementation option is to deploy an equal number of transceivers and antennas,
and then simply turn on the transceivers corresponding to the selected antennas while turning off the rest.  
This implementation is illustrated in Fig.~\ref{fig:system_model} and is more flexible 
as the number of active antennas can be variable. 
With power switches,
the reduction in system complexity relies on simpler implementation of the baseband signal processing
due to a reduced number of active antennas and RF chains.
However, with a variable number of active antennas,
the antenna selection algorithms will add extra complexity,
e.g., in making a decision on the optimal number of antennas.

%As a result of antenna selection, we can reduce the number of RF transceivers in two optional ways: 
%1) reducing the actual number of RF transceivers that are deployed at the base station,
%and 2) turning off the RF transceivers that are not needed at a certain time interval.
%In the first option,
%the selected $N$ antennas have to be connected with the $N$ RF transceivers through an RF switch.
%The implementation of RF switch can be highly complex, as both $M$ and $N$ become large.
%Additionally, it introduces extra signal power attenuation and hardware cost.
%With massive MIMO, it may be more feasible to keep the same number of RF transceivers as the antennas
%so that RF switching can be avoided,
%and we simply turn on the transceivers associated with the selected antennas
%while turning off the others.
%The second option is also more flexible in case that more transceivers are needed to achieve required performance, 
%e.g., when we switch between precoding/detection schemes, 
%or more users are served on the same time-frequency resource.

Antenna selection has been widely studied for conventional MIMO,
see for example \cite{AntSel_Molisch, Molisch_Capacity_AntSel2005, Sanayei_AntSel, Sanayei_Capacity_MIMO_antSel_2007, Gucluoglu_AntSel_2008,
Sadek_Active_MUMIMO_AntSel_2007, TX_AntSel_MUMIMO_2007, Mohaisen_AntSel_DPC_2009, Lin_TX_AntSel_MUMIMO_2012}.
%such as in \cite{AntSel_Molisch, Molisch_Capacity_AntSel2005, Sanayei_AntSel, Sanayei_Capacity_MIMO_antSel_2007, Gucluoglu_AntSel_2008} 
%for single-user MIMO (SU-MIMO),
%and in \cite{Sadek_Active_MUMIMO_AntSel_2007, TX_AntSel_MUMIMO_2007, Mohaisen_AntSel_DPC_2009, Lin_TX_AntSel_MUMIMO_2012} for multi-user (MU-MIMO).
However, to the best of our knowledge, 
there are only few studies on antenna selection for massive MIMO available.
In \cite{Dong_MaMIMO_AntSel_60GHz_2011},
antenna selection in massive MIMO was addressed for short-range wireless communications at 60~GHz. 
In \cite{BAN2013}, a simulation study using the Kronecker channel model \cite{Kermoal_Kronecker_Model_2002} showed that significantly higher performance can be achieved with antenna selection than without.
In \cite{Gkizeli2014}, antenna selection for maximizing signal-to-noise ratio (SNR) was studied,
and \cite{Benmimoune2015} considered antenna selection jointly with user scheduling for massive MIMO.
The authors in \cite{Kataoka2014} evaluated the characteristics of interference rejection with antenna sector selection in massive MIMO,
based on measured channels in the 2~GHz band with 96 antenna elements.
In the conference paper \cite{Ant_Sel_Globecom}, we presented preliminary results on antenna selection in measured massive MIMO channels. 
The current paper extends \cite{Ant_Sel_Globecom} by studying in more depth how many RF chains that can be switched off while achieving a required performance, by considering more scenarios and propagation conditions, 
and by performing comparisons with suboptimal precoding schemes.  
 
\section{Approach}\label{sec:approach}

The aim of this paper is to obtain a deeper insight into how antenna selection in massive MIMO performs in real propagation channels. Specifically, we focus on how the number of users, the separation of users, and propagation conditions like LOS and NLOS
affect the performance of antenna selection.
%\todo[inline]{Do we need to motivate why we study the effect of the number of users, the separation of users and LOS/NLOS here?  
%For example, when the user channels are distinct and there are many users, antenna selection would be less effective. //Xiang}
Although the large power variation across antennas remains when averaging over frequencies,
the effectiveness of the antenna selection can be reduced if the user channels to the base station are very distinct.
This happens when users are located far apart and when many users are served.
It is demonstrated that in the ``worst case'', 
adaptive antenna selection does not perform significantly better than random selection, 
but in many cases adaptive selection substantially improves over random selection.
All investigations use the measured channel data at 2.6~GHz described above, 
obtained with linear and cylindrical arrays.
 
In terms of algorithms, we select the set of active
antennas that maximizes the downlink capacity.
To find the optimal set, an exhaustive search can be used; 
however this is infeasible for massive MIMO in practice due to the huge number of possible selection alternatives.
A number of antenna selection algorithms with lower complexity, 
notably greedy selection, have been proposed for conventional MIMO,
and many of them can be applied to massive MIMO.
We first examine a near-optimal scheme that uses convex optimization \cite{AntSel_convex_Dua2006, Transmit_AntSel_Convex_2012, Ant_Sel_Globecom}.
We then consider a very simple selection scheme that is based only on measurements of the received power at each antenna.
Generally, this power-based selection scheme underperforms the convex-optimization based scheme
that considers not only the received power but also the correlation between antenna channels.
Yet, experiments with measured data show that the power-based scheme performs fairly well. 
This is so because the power variations over the array can be considerable in massive MIMO. 

\section{System Description and Antenna Selection Schemes}\label{sec:system_model}

We first establish the system model that will be used in the rest of the paper. 
We also formally state the problem of antenna selection for downlink capacity maximization, 
and introduce the two selection schemes: a near-optimal scheme relying on convex optimization,
and a simple scheme using only received signal power measurements.

\subsection{System Model and Sum-Capacity}

\begin{figure}
  \centering
  \includegraphics[width=0.48\textwidth]{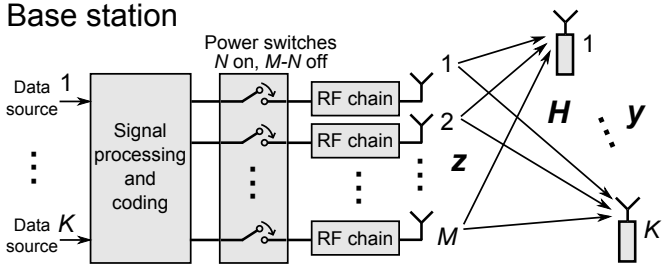} %- for double-column
  \caption{Multi-user MIMO system with transmit antenna selection. 
The base station has $M$ available antennas and $N$ active RF chains, 
and serves $K$ single-antenna users in the same time-frequency resource.
The switches indicate that entire RF chains are being switched on or off.}
  \label{fig:system_model}
\end{figure}

We consider a single-cell multi-user MIMO-OFDM system with $L$ subcarriers in the downlink.
As shown in Fig.~\ref{fig:system_model}, 
the base station has $M$ antennas, and each antenna has an associated transceiver chain.
With $N$ antennas being selected, the $N$ corresponding transceivers are switched on, while the other $M\!-\!N$ are switched off.
%The RF chains are connected to $N$ selected antennas, making $M\!-\!N$ antennas not connected.
This base station with $N$ active antennas and transceivers serves $K$ single-antenna users 
%($K\!\leq\!N\!\leq\!M$) 
in the same time-frequency resource.
With massive MIMO, we assume $M\!\gg\!K$ and allow $N$ to be in the range from $K$ to $M$.

The model for the downlink channel is
\begin{equation}\label{eq:signal_model}
	\vec{y}_\ell=\sqrt{\rho K}\vec{H}^{\left(N\right)}_\ell\vec{z}_\ell+\vec{n}_\ell,
\end{equation}
where $\vec{H}^{\left(N\right)}_\ell$ is a $K\!\times\!N$ channel matrix at subcarrier $\ell$,
and the superscript $\left(N\right)$ indicates that antenna selection has been performed, i.e., the $N$ columns of $\vec{H}^{\left(N\right)}_\ell$ 
are selected from the $K\!\times\!M$   full channel matrix $\vec{H}_\ell$.
Normalization is performed such that the elements of $\vec{H}^{\left(N\right)}_\ell$ have unit  energy, averaged over  all $L$ subcarriers,
$M$ antennas and $K$ users, see  \cite{Gao_MassiveMIMO_Real_Channel_2014} for more details.
Then $\vec{z}_\ell$ is the $N\!\times\!1$ transmit vector across the $N$ selected antennas, 
and satisfies $\mathbb{E}\left\{\|\vec{z}_\ell\|^2\right\}\!=\!1$,
$\vec{y}_\ell$ is the received vector at the $K$ users, 
and $\vec{n}_\ell$ is a noise vector with i.i.d.\ complex Gaussian, $CN(0,1)$, elements. 
The factor $\rho K$ represents the transmit power. With the conventions used in this paper,
the transmit power per user is fixed. Hence, the total transmit power increases 
with $K$ but is independent of $N$.
%\textcolor{blue}{Here we do not scale the transmit power by $1/N$ as in 
%\cite{Gao_MassiveMIMO_Real_Channel_2014} and \cite{Jose_ICC_2014}.
%The reason is to facilitate a fair performance comparison when the number of active antennas is varied.}
The parameter $\rho$ represents the normalized transmit SNR per user. 
With random antenna selection, the average per-user received SNR would be $\rho N$,\footnote{The received SNRs at the users
in general depend on precoding scheme and the channel conditions.
For example, in the single-user case, the received SNR is $\rho N$.
In the multi-user case with zero-forcing precoding, the per-user received SNR is
$\rho K / \Tr\left\{\left(\vec{H}_\ell^{\left(N\right)}\left(\vec{H}_\ell^{\left(N\right)}\right)^H\right)^{-1}\right\}$,
where $\Tr\left\{\cdot\right\}$ represents the trace of a matrix.
When the user channels are orthogonal, %(called ``favorable'' propagation conditions \cite{Ngo2014_favorable}),
$\vec{H}_\ell^{\left(N\right)}\left(\vec{H}_\ell^{\left(N\right)}\right)^H$ is diagonal,
and the average per-user received SNR reaches the upper bound given by the single-user case, i.e., $\rho N$.
Under ``favorable'' propagation conditions \cite{Ngo2014_favorable},
the user channels becomes orthogonal when the number of base station antennas grows,
thus the average received SNR approaches this upper bound.
%To be exact, 
%with random antenna selection,
%the average per-user received SNR is smaller or equal to $\rho N$,
%depending on channel condition and used precoding scheme.
%In ``favorable'' propagation conditions that
%the user channels are orthogonal \cite{Ngo2014_favorable}, i.e., 
%when $\vec{H}_\ell^{\left(N\right)}\left(\vec{H}_\ell^{\left(N\right)}\right)^H$ is diagonal,
%the average per-user SNR is equal to $\rho N$, 
%for both dirty-paper coding (DPC) and zero-forcing (ZF) precoding.
%When user channels are not completely orthogonal and inter-user interference exists, 
%the average per-user SNR is smaller than $\rho N$,
%and the SNR using DPC is higher than using ZF precoding.
}
which increases with the number of selected antennas $N$ due to the increased array gain. 
%It is then straightforward to know how many RF transceivers can be switched off,
%i.e., to what extent the array gain can be afforded to lose,
%while required performance can still be achieved.
When the number of users $K$ varies, the average per-user received SNR is constant, and so is the average per-user rate (disregarding interference),
if a fixed number $N$ of RF transceivers are switched on.
With adaptive antenna selection,
the received SNRs are expected to be higher than those with random antenna selection,
since the ``best'' antennas are selected.

To avoid favoring users that have a better average channel,    
we normalize the channel matrix to remove the effects of the pathloss and the large-scale fading while retaining  
the effects of the small-scale fading.  
Specifically, when the users are far apart,
we normalize the channel matrix according to Normalization 1 in \cite{Gao_MassiveMIMO_Real_Channel_2014}, 
and when the users are closely located, Normalization 2 in \cite{Gao_MassiveMIMO_Real_Channel_2014} is applied.
%This is because users with similar required power levels are usually grouped and served simultaneously 
%so that each user can have a relatively good data rate. 
%\todo[inline]{Ove: Is there a reference for this grouping of users with similar powers? \\
%Xiang: I checked some references that trade off sum-rate maximization and fairness. 
%Can we say that for fairness we assume users with similar required power levels are grouped together?}
%This is because that the large attenuation imbalance will make the antenna selection favor the the user channels with higher power,
%while the user channels with lower power have smaller influence on the antenna selection decision.
%Ideally, in order to study how spatial structure of multi-user MIMO channels, rather than attenuation imbalance, affects antenna selection,
%we would like to group the users with similar channel gain together.
%However, due to limited number of measurement positions, we do not have such a case when users are far apart.
%Therefore, we normalize the large power imbalance between users,
%while maintaining the spatial property of the channels from each user to the base station.
However, importantly, we do not normalize the channel variations per base station antenna, 
since these variations are critical for the antenna selection.

With the defined signal model, 
the downlink sum-capacity at subcarrier $\ell$ is given by \cite{dpc_Vishwanath2003}:
\begin{equation}\label{eq:dpc_capacity_1}
	C_{\mathrm{DPC},\ell}=\max_{\vec{P}_\ell}\log_2\det\left(\vec{I}+\rho K\left(\vec{H}^{\left(N\right)}_\ell\right)^H\vec{P}_\ell\vec{H}^{\left(N\right)}_\ell\right),
\end{equation}
which is achieved using   dirty-paper coding (DPC) \cite{Costa1983_DPC}.
In (\ref{eq:dpc_capacity_1}), $\vec{P}_\ell$ is a diagonal power allocation matrix  with $P_{\ell,i},i\!=\!1,2,...,K$ on its diagonal.
Also, in (\ref{eq:dpc_capacity_1}), the optimization is performed subject to the total power constraint that $\sum_{i\!=\!1}^K{P_{\ell,i}}\!=\!1$. 
This optimization problem is convex and can be solved, for example, by using the sum-power iterative waterfilling algorithm in \cite{dpc_wf_Jindal2005}.

DPC is highly complex to implement in practice.
However, there are suboptimal linear precoding schemes, such as zero-forcing (ZF) precoding %and maximum-ratio (MR) precoding
that is much less complex and performs fairly well for massive MIMO \cite{emil-myth, Xiang_Asilomar}.
The sum-rate achieved by ZF precoding is \cite{zf_Wiesel2008}
\begin{equation}\label{eq:zf_rate}
	C_{\mathrm{ZF},\ell}=\max\limits_{\vec{Q}_\ell}\sum_{i=1}^K{\log_2\left(1+\rho K Q_{\ell,i}\right)},
\end{equation}
where $Q_{\ell,i}$ represent received SNRs of the different users and
the maximization is performed subject to the total power constraint
\begin{equation}\label{eq:zf_constraint}
	\sum_{i=1}^K{Q_{\ell,i}\left[\left(\vec{H}^{\left(N\right)}_\ell\left(\vec{H}^{\left(N\right)}_\ell\right)^H\right)^{-1}\right]_{i,i}}=1.
\end{equation}
In (\ref{eq:zf_rate}) and (\ref{eq:zf_constraint}), 
$\vec{Q}_\ell$ is a diagonal matrix with $Q_{\ell,i},i\!=\!1,2,...,K$ on its diagonal,
and $\left[\cdot\right]_{i,i}$ indicates the $i$-th diagonal element of a matrix.
The diagonal elements of $\left( \vec{H}^{\left(N\right)}_\ell\left(\vec{H}^{\left(N\right)}_\ell\right)^H\right)^{-1}$ represent the power penalty of nulling out interference.
The optimization in (\ref{eq:zf_rate}) can be solved using the standard waterfilling algorithm \cite{Cover1991}.

We choose to base the antenna selection algorithms on  the DPC sum-capacity. However,
performance of the resulting selection will be evaluated in terms of ZF sum-rate too, in relevant cases.
Note that different antenna combinations can be optimal on different subcarriers. However, 
in a practical MIMO-OFDM system, the same antennas need be selected for all subcarriers. 
Therefore, our algorithms will find a set of $N$ antennas that maximizes the DPC capacity averaged over all $L$ subcarriers.

To select the $N$ columns from the full MIMO matrix $\vec{H}_\ell$,
we introduce an $M\!\times\!M$ diagonal matrix $\vec{\Delta}$, with binary diagonal elements
\begin{equation}\label{eq:delta_matrix}
	\Delta_i=
	\begin{cases}
	1,\ \mathrm{selected} \\
	0,\ \mathrm{otherwise},
	\end{cases}
\end{equation}
indicating whether the $i$th antenna is selected, and satisfying $\sum_{i\!=\!1}^M{\Delta_i}\!=\!N$.
Using Sylvester's determinant identity, 
$\det{\left(\vec{I}\!+\!\vec{A}\vec{B}\right)}\!=\!\det{\left(\vec{I}\!+\!\vec{B}\vec{A}\right)}$, 
we can write the DPC sum-capacity in (\ref{eq:dpc_capacity_1}) in terms of 
  $\vec{\Delta}$  as
\begin{eqnarray}\label{eq:dpc_capacity_2}
	C_{\mathrm{DPC},\ell}\!&=&\!\max_{\vec{P}_\ell}\log_2\det\left(\vec{I}+\rho K\vec{P}_\ell\vec{H}^{\left(N\right)}_\ell\left(\vec{H}^{\left(N\right)}_\ell\right)^H\right)\nonumber \\
	\!&=&\!\max_{\vec{P}_\ell}\log_2\det\left(\vec{I}+\rho K\vec{P}_\ell\vec{H}_\ell\vec{\Delta}\vec{H}_\ell^H\right),
\end{eqnarray}
subject to $\sum_{i\!=\!1}^K{P_{\ell,i}}\!=\!1$.
%For the corresponding ZF sum-rate, expressed in terms of $\vec{\Delta}$, 
%the total power constraint in (\ref{eq:zf_constraint}) becomes
%\begin{equation}\label{eq:zf_constraint_2}
	%\sum_{i=1}^K{Q_{\ell,i}\left[\left(\vec{H}_\ell\vec{\Delta}\vec{H}_\ell^H\right)^{-1}\right]_{i,i}}=1.
%\end{equation}
The optimal $\vec{\Delta}$ (common to all subcarriers) is found by maximizing the average DPC capacity,
\begin{equation}\label{eq:delta_opt}
	\vec{\Delta}_\mathrm{opt}\!=\!\underset{\vec{\Delta}}{\arg\max}\ \frac{1}{L}\sum_{\ell=1}^L{\left\{\log_2\det\left(\vec{I}\!+\!\rho K\vec{P}_\ell\vec{H}_\ell\vec{\Delta}\vec{H}_\ell^H\right)\right\}}.
\end{equation}
With the resulting antenna selection, we have the corresponding ZF sum-rate
\begin{equation}\label{eq:zf_rate_2}
	C_{\mathrm{ZF},\ell}=\max\limits_{\vec{Q}_\ell}\sum_{i=1}^K{\log_2\left(1+\rho K Q_{\ell,i}\right)},
\end{equation}
subject to
\begin{equation}\label{eq:zf_constraint_2}
	\sum_{i=1}^K{Q_{\ell,i}\left[\left(\vec{H}_\ell\vec{\Delta}_\mathrm{opt}\vec{H}_\ell^H\right)^{-1}\right]_{i,i}}=1.
\end{equation}
Note that $\vec{\Delta}_\mathrm{opt}$ may not be optimal for ZF.
Despite this, the ZF sum-rate indicates the antenna selection performance when using a more practical precoding scheme than DPC.

As discussed in Sec.~\ref{sec:intro}, exhaustive search of all possible combinations of $N$ antennas will certainly give us the optimal $\vec{\Delta}$,
however, it is extremely complex and infeasible for massive MIMO.
We next introduce two practical selection schemes that will be used in our performance study 
in Sec.~\ref{subsec:convex} and Sec.~\ref{subsec:power}.

\subsection{Antenna Selection Using Convex Optimization}\label{subsec:convex}

Here we assume that the base station has perfect channel state information (CSI).
The near-optimal selection scheme using convex optimization was introduced and used in \cite{Ant_Sel_Globecom}.
We give a brief description in the following.
As can be seen in (\ref{eq:delta_opt}),
to maximize the average DPC capacity over subcarriers,
we need to optimize over both $\vec{\Delta}$ and $\vec{P}_\ell$.
This is a difficult task and we therefore divide the optimization into two steps:
1) we assume equal power allocation among the users, 
i.e., $P_{\ell,i}\!=\!1/K$, and select the $N$ antennas that maximize the average capacity;
2) with the selected $N$ antennas, we optimize over $\vec{P}_\ell$ on each subcarrier, 
and thus obtain the maximum average capacity for the case of $N$ antennas.
Although this simplification does not ensure that we find the global optimum,
it gives us a lower bound on the performance we can achieve by using adaptive antenna selection.

In Step~1, the optimization problem of antenna selection can be formulated as
\begin{equation}\label{eq:original_problem}
	\begin{aligned}
		& \text{maximize} 
		& & \frac{1}{L}\sum_{\ell=1}^L\left\{\log_2\det\left(\vec{I}+\rho\vec{H}_\ell\vec{\Delta}\vec{H}_\ell^H\right)\right\}, \\
		& \text{subject to} 
		& & \Delta_i\in\left\{0,1\right\} \\
		%& & \vec{\Delta}_i=0\:\text{or}\:1 \\
		& & & \sum_{i=1}^{M}{\Delta_i}=N.
	\end{aligned}
\end{equation}
The objective function  is concave in $\vec{\Delta}$ \cite{Convex_Boyd2004}.
However, the variables $\Delta_i$ are binary integer variables, which makes the optimization problem NP-hard.
In order to solve this optimization problem, as in   \cite{AntSel_convex_Dua2006,Transmit_AntSel_Convex_2012}, 
we relax the constraint that each $\Delta_i$ must be binary integer to the weaker constraint that $0\!\leq\!\Delta_i\!\leq\!1$.
The original problem thus becomes a convex optimization problem solvable in polynomial time. 
This relaxation yields a  solution with non-integral values of $\Delta_i$. 
From the relaxed solution, the $N$ largest $\Delta_i$  are selected, and their indices represent the selected antennas.
As discussed in \cite{Ant_Sel_Globecom,AntSel_convex_Dua2006,Transmit_AntSel_Convex_2012},
the relaxation gives near-optimal results,
except for when we select a very small number of antennas, 
i.e., $N\!\ll\!M$.
In a massive MIMO system, $N$ should be relatively large and
therefore we believe that the relaxation method  is technically sound.
%
%With the largest $N$ $\Delta_i$'s rounded up to 1, and the rest set to 0, 
%the near-optimal antenna selection matrix is thus obtained,
%which we denote $\vec{\Delta}_\mathrm{sub\!-\!opt}^\mathrm{conv}$.
%%$\vec{\Delta}_{\mathrm{near}\tilde{-}\mathrm{opt}}$ is therefore obtained.
%Finally, we maximize the average capacity by optimizing transmit power allocation among users at each subcarrier.
%%the average capacity is maximized by optimizing power allocation among users at each subscarrier.
%The obtained average DPC capacity can be written as
%\begin{equation}\label{eq:power_allocation}
%	C_\mathrm{DPC}^\mathrm{conv}=\mathbb{E}_\ell\left\{\log_2\det\left(\vec{I}+\rho K\vec{P}_\ell\vec{H}_\ell\vec{\Delta}_\mathrm{sub\!-\!opt}^\mathrm{conv}\vec{H}_\ell^H\right)\right\}.
%\end{equation}
%Along with the average ZF sum-rate obtained from the $N$ selected antennas, denoted as $C_\mathrm{ZF}^\mathrm{conv}$,
%we evaluate the performance of antenna selection in Sec.~\ref{sec:results}.

\subsection{Antenna Selection Based on Received Power}\label{subsec:power}

Using only the received power per antenna as the basis for antenna selection results in a very simple scheme.
We select the $N$ antennas that have the highest received power from all $K$ users, 
averaged over all $L$ subcarriers.
As compared to the convex-optimization based scheme, 
the power-based scheme has very low complexity.
By only measuring the received power at each antenna branch in the uplink (exploiting channel reciprocity),
we can make a decision on the antenna selection for the downlink 
%without estimated CSI and complex signal processing.
before any CSI estimation is performed and without complex signal processing.
As discussed in Sec.~\ref{sec:intro}, this simple selection scheme generally shows worse performance than the convex-optimization scheme.
However, in situations when all antenna channels have relatively low correlation, 
e.g., in NLOS scenarios with rich scattering, 
the power-based selection scheme may become near-optimal.
We compare the performance obtained through the two selection schemes with  measured channels,
for different propagation scenarios, in Sec.~\ref{sec:results}.

\section{Measured channels}\label{sec:measurements}

The channel measurements used in this paper were first reported in \cite{Xiang_Asilomar,Gao_MassiveMIMO_Real_Channel_2014}.
Here, we give a brief summary.

\begin{figure}
  \centering
  \includegraphics[width=0.48\textwidth]{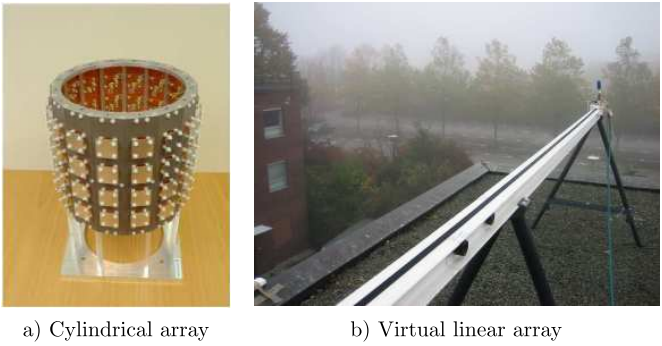} %- for double-column
  \caption{Two large antenna arrays at the base station side:
  a) a cylindrical array with 64 dual-polarized patch antenna elements, giving 128 ports in total,
  and b) a virtual linear array with 128 vertically-polarized omni-directional antennas.}
  \label{fig:bs_array}
\end{figure}

Measurements were taken over bandwidth of 50~MHz on the 2.6~GHz band,
using two different large antenna arrays (cylindrical and linear) at the base station. 
Both arrays contain 128 antenna elements and have an adjacent element spacing of half a wavelength. 
Fig.~\ref{fig:bs_array}(a) shows the cylindrical array,
comprising 16 dual-polarized directional patch antennas in each circle with 4 such circles stacked on top of each other, 
giving a total of 128 antenna ports. 
This array is physically compact with physical dimensions (both diameter and height) of about 30~cm.
Fig.~\ref{fig:bs_array}(b) shows the virtual linear array with a vertically-polarized omni-directional antenna moving between 128 equidistant positions, along a rail.
The linear array is 7.4~m long, which is more than 20 times the size of the cylindrical array.
In both measurement campaigns, an omni-directional antenna with vertical polarization was used at the user side.

\begin{figure}
  \centering
  \includegraphics[width=0.48\textwidth]{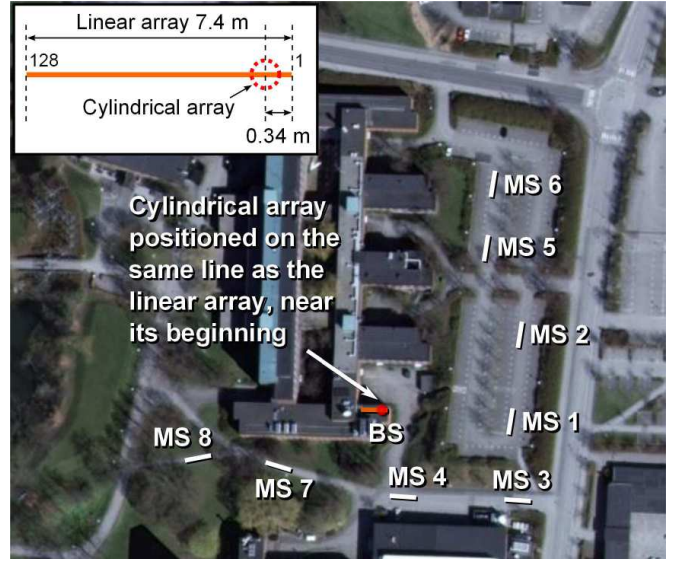} %- for double-column
  \caption{Overview of the measurement area at the Faculty of Engineering (LTH) campus at Lund University in Sweden. 
	The two base station antenna arrays were placed on the same roof of the E-building. 
  At the user side, eight sites (MS~1-8) around the E-building were measured.}
  \label{fig:MAP_small}
\end{figure}

All measurements were carried out outdoors at the E-building of
the Faculty of Engineering (LTH) of Lund University in Sweden.
Fig.~\ref{fig:MAP_small} shows an overview of the semi-urban measurement area.
The two base station antenna arrays were placed on the same roof of the E-building during their respective measurement campaigns. 
More precisely, the cylindrical array was positioned on the same line as the linear array, near its beginning,
and was for practical reasons mounted about 25~cm higher than the linear array.
At the user side, the omni-directional antenna was moved between eight measurement sites (MS~1-8) around the E-building, emulating 
single-antenna users. 
Among these eight sites, three (MS~1-3) have LOS conditions, 
and four (MS~5-8) have NLOS conditions, while one (MS~4) has LOS for the cylindrical array, 
whereas the LOS component is blocked by the roof edge for the linear array.
Despite this, MS~4 still has LOS characteristic for the linear array, where one or two dominating
multipath components due to diffraction at the roof edge cause a relatively high Ricean K-factor \cite{Molisch2005}.

The investigations in \cite{Xiang_Asilomar} and \cite{Gao_MassiveMIMO_Real_Channel_2014} showed that 
the linear array achieves higher average sum-rates than the cylindrical array, 
if we randomly select the same number of antennas on both arrays.
The reason is that the linear array has very high angular resolution due to its large aperture, 
which helps to spatially separate the users,
especially when users are closely located at the same measurement site.
The cylindrical array has smaller aperture thus lower angular resolution, 
and due to its circular arrangement some antennas may face the ``wrong'' directions and contribute little.
With a vertically-polarized antenna at the user side,
the dual-polarization arrangement at the base station also degrades the performance of the cylindrical array,
but only to a certain extent.
In the measured channels, the received power ratios of the vertically-polarized and horizontally-polarized antenna ports 
are approximately log-normal distributed,
with a mean value of 2.2~dB and a standard deviation of 8~dB in the dB domain.
Note that the above investigations and comparisons are based on the spatial structure of the two arrays,
when the two arrays have equal average channel gain due to the performed normalization, as discussed in Sec.~\ref{sec:system_model}.
In reality, however, the cylindrical array may perform better than what we have seen, 
when the antenna gains of the patch elements are taken into account. 
Also, antennas at the user side are usually dual-polarized in reality, 
making both polarizations at the base station useful for user separation \cite{Jose_ICC_2014}.
However, it is not a priori clear which array that performs better, if we adaptively select antennas. 
We investigate this matter in the next section.
 
\section{Performance Results in Measured Channels}\label{sec:results}

With the obtained channel data, 
we apply antenna selection, as described in Sec.~\ref{sec:system_model},
for both arrays, in different propagation scenarios and for different number of users.
Note that all results are obtained from the measured channels.
First we focus on the convex-optimization scheme, since it gives us near-optimal results. 
We investigate how much we can gain by performing antenna selection, as compared to random selection,
and how many RF transceivers we can switch off while maintaining 90\% of full MIMO performance.
Then we move to the simple selection scheme based on only received power measurements, 
and compare the corresponding performance with that of the convex-optimization scheme.

The parameter setting for evaluating antenna selection performance is as follows.
We have $M\!=\!128$ antennas at the base station, among which we select the $N$ that works ``best'' across all $L\!=\!161$ subcarriers, depending on which antenna selection scheme is used.
We perform antenna selection for $N$ growing from $K$ to 128. When $N\!=\!128$, we have the full MIMO performance.
We study cases where the number of users, $K$, is 4, 16 and 40, respectively.
In all cases, we set $\rho\!=\!-5$~dB, 
so that in the interference-free case and with random antenna selection the average per-user rate is in the range of 1.2-5.4~bps/Hz,
as $N$ grows to 128 and the array gain increases accordingly.
The range of the per-user rate does not depend on the number of users, since we maintain the same transmit power per user,
as discussed in Sec.~\ref{sec:system_model}.
Next, we present and discuss the results.

\subsection{Performance of Convex-Optimization Selection Scheme}\label{sec:results_convex}

To investigate the effectiveness of antenna selection in different propagation scenarios, 
we first focus on the case of four users ($K\!=\!4$), 
which is the number of simultaneous users supported in multi-user MIMO transmission in LTE \cite{3GPP_LTE_A}.
Then we increase the number of users to sixteen ($K\!=\!16$) and forty ($K\!=\!40$),
and investigate the corresponding performance,
as massive MIMO is capable of serving more users.

\subsubsection{\textbf{Four users}, $K\!=\!4$}
Combining user separation and LOS/NLOS condition,
here we choose two reference scenarios to study, in which
the four users are
\begin{itemize}
	\item close to each other (1.5-2~m spacing), all at MS~2, having LOS conditions to the base station,
	\item well separated (larger than 10~m spacing), at MS~5-8, respectively, all having NLOS conditions to the base station.
\end{itemize}
We expect more effective use of antenna selection in the first scenario, 
since the channels are less frequency-selective and less distinct to different users.
%since the spatial diversity on the base station array tends to be more similar for the co-located users, 
%and the channel with LOS condition should be more frequency-flat.
In the second scenario, there is higher frequency selectivity due to the NLOS conditions.
Also, the users are more widely separated.
Hence, it is expected that the combination of antennas that are optimal for a given user on a given subcarrier 
differs between the users and the subcarriers. 
Antenna selection, where the same antennas
are used for all subcarriers and all users, will therefore be less effective in this scenario.
%higher frequency selectivity due to NLOS condition, and wider separation of users, 
%may cause different antenna combinations being optimal at different subcarriers and for different users,
%which reduces effectiveness of antenna selection.

\begin{figure}
  \centering
  \includegraphics[width=0.48\textwidth]{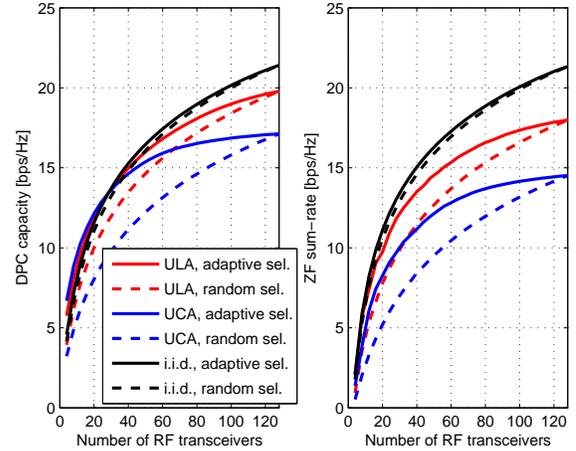}
  \caption{Performance of adaptive antenna selection using the convex-optimization scheme,
	as compared to performance of random selection,
  in the LOS scenario where four users are closely located at MS~2.
	``ULA'' and ``UCA'' stand for uniform linear array and uniform cylindrical array, respectively.
	%With 40 active RF transceivers, the performance gain by adaptive antenna selection is 11-18\% for the linear array, and more than 30\% for the cylindrical array, compared to random selection.
  %For the linear array, 70 transceivers reach 90\% of full MIMO performance, with both DPC and ZF.
  %For the cylindrical array, 50 or 60 transceivers achieve 90\% of full MIMO performance with DPC or ZF.
  %In this scenario, a substantial number of transceivers can be switched off without losing substantial performance.
  }
  \label{fig:ant_sel_gain_los_b_fix_tx_pow}
\end{figure}

\begin{figure}
  \centering
  \includegraphics[width=0.48\textwidth]{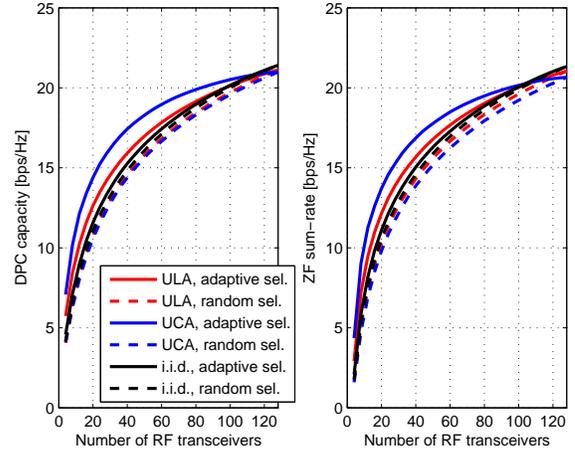} %for double-column
  \caption{Performance of adaptive antenna selection using the convex-optimization scheme,
	as compared to performance of random selection,
  in the NLOS scenario where four users are well separated at MS~5-8.
  ``ULA'' and ``UCA'' stand for uniform linear array and uniform cylindrical array, respectively.
	%With 40 active RF transceivers, the performance gain by the adaptive antenna selection is 10\% and 20\% for the linear and cylindrical array, respectively, compared to random selection.
	%To reach 90\% of full MIMO performance, 80 and 60 transceivers are needed for the linear and cylindrical array, respectively,
	%with both DPC and ZF.
  %For the linear array, 80 RF transceivers reach 90\% of full MIMO performance, with both DPC and ZF precoding,
  %while 60 RF transceivers are needed for the cylindrical array.
  %Still, a large number of RF transceivers can be switched off in this scenario.
	}
  \label{fig:ant_sel_gain_nlos_sp_fix_tx_pow}
\end{figure}

The resulting DPC capacities and ZF sum-rates by performing antenna selection are shown
in Fig.~\ref{fig:ant_sel_gain_los_b_fix_tx_pow} and Fig.~\ref{fig:ant_sel_gain_nlos_sp_fix_tx_pow}, 
for the two scenarios, respectively.
As a reference, we also show the antenna selection performance in i.i.d. Rayleigh channels.
We can see that in i.i.d. Rayleigh channels the performance gain by applying adaptive antenna selection is very small,
both in DPC capacity and ZF sum-rate, 
as compared to the average performance obtained from random selection of antenna combinations.
This indicates that antenna selection is quite ineffective in i.i.d. Rayleigh channels,
since all antennas are equally good, as discussed in Sec.~\ref{sec:background}.

In Fig.~\ref{fig:ant_sel_gain_los_b_fix_tx_pow} where the four users are closely located with LOS,
the measured channels provide significantly larger gain when performing antenna selection,
for both arrays.
With 40 RF transceivers, i.e., 10 times the number of users,
for the linear array with antenna selection, the DPC capacity and ZF sum-rate increase by 11\% and 18\%,
as compared to the performance of random selection.
For the cylindrical array the gain is even higher, more than 30\%, both in DPC capacity and ZF sum-rate.
We can also see the performance loss when switching off RF transceivers.
For the linear array, about 70 RF transceivers can achieve 90\% of the full MIMO performance, with both DPC and ZF precoding.
For the cylindrical array, only about 50 and 60 are needed, with DPC and ZF, respectively,
thus more than half of the RF transceivers can be switched off.
This can be explained that in this particular scenario many antennas on the cylindrical array do not ``see'' the users.
%These observations indicate that we can trade sum-rates with the number of RF transceivers.
%In this particular scenario, since some antennas significantly outperform the others, 
%we can substantially reduce the number of RF transceivers without losing too much in performance.
%This scenario allows for very effective use of antenna selection. 

We next consider the scenario where the four users have NLOS conditions and are well separated.
As shown in Fig.~\ref{fig:ant_sel_gain_nlos_sp_fix_tx_pow},
the performance gain when performing adaptive antenna selection drops in the measured channels,
compared to the previous scenario.
The higher frequency selectivity due to NLOS conditions and the wider separation of users indeed reduce 
the effectiveness of antenna selection to some extent.
Despite this, we still observe some gains in the measured channels, as compared to i.i.d. Rayleigh channels.
At 40 RF transceivers, using adaptive antenna selection,
we increase both the DPC capacity and ZF sum-rate by 10\% for the linear array,
and 20\% for the cylindrical array,
as compared to random selection.
Correspondingly, to achieve 90\% of the full MIMO performance,
a slightly higher number of RF transceivers are needed in this scenario.
For the linear array, we need 80 RF transceivers,
while for the cylindrical array, we need around 60, with both DPC and ZF.
Still, a large number of RF transceivers can be switched off in this scenario.

From Fig.~\ref{fig:ant_sel_gain_nlos_sp_fix_tx_pow}, 
another important observation in this scenario is that
by using adaptive antenna selection, 
the measured channels with both arrays achieve higher performance than i.i.d. Rayleigh channels,
except for when nearly all transceivers are active. 
With random selection, i.i.d. Rayleigh channels give better average performance than the measured channels,
also reported in \cite{Gao_MassiveMIMO_Real_Channel_2014}.
However, by exploiting the large number of spatial degrees of freedom in the measured channels through adaptive antenna selection, 
the transmit energy is fed to those ``best'' antennas with relatively high channel gains and 
relatively low correlation between each other,
thus performance increases.
Compare with the LOS scenario in Fig.~\ref{fig:ant_sel_gain_los_b_fix_tx_pow};
there the measured channels cannot outperform i.i.d. Rayleigh channels. 
The reason is that the spatial separation is particularly difficult for the closely-spaced users under LOS,
therefore, even with adaptive antenna selection, 
the performance in the measured channels cannot surpass that of i.i.d. channels.
In the NLOS scenario, however, the channel correlation between the well-separated users is relatively low, as in i.i.d. Rayleigh channels,
hence, by selecting the antennas with relatively high channel gains, 
the measured channels outperform i.i.d. Rayleigh channels.
Especially for the cylindrical array, the performance is significantly improved,
and is higher than that of the linear array, for a large range of active transceiver numbers.
Thus,
adaptive antenna selection provides an opportunity for the cylindrical array to achieve higher performance.
Taking practical deployments into consideration, this small and compact array is preferable to the physically large linear array.

From the above evaluation, adaptive antenna selection is effective in both scenarios.
We gain significantly not only in DPC sum-capacity, which is our antenna selection criterion, 
but also in the sum-rate obtained by more practical ZF precoding.
With four users, we can switch off 50-60 RF transceivers for the linear array, 
and 70-80 RF transceivers for the cylindrical array, while only losing 10\% of full MIMO performance.
However, if more users are served, 
more antennas and transceivers are required to spatially separate users.
%In other words, the effectiveness of antenna selection may reduce, 
%as the ``massive'' spatial diversity averages out over a larger number of users, 
%especially when they are well distributed around the base station. 
In this case, since more antennas are contributing, can we still switch off many RF transceivers?
Next we increase the number of users to sixteen ($K\!=\!16$) and forty ($K\!=\!40$), 
and investigate the corresponding antenna selection performance.

\subsubsection{\textbf{Sixteen and forty users}, $K\!=\!16$ and $K\!=\!40$}
Here we have sixteen or forty users distributed at MS~1-8.
When $K\!=\!16$, two users are at the same site with 5~m spacing.
When $K\!=\!40$, five users placed at each site with 0.5-2~m spacing.
Users at different sites are spaced more than 10~m apart.
Among these users, half have LOS conditions and half have NLOS conditions.

\begin{figure}
  \centering
  \includegraphics[width=0.48\textwidth]{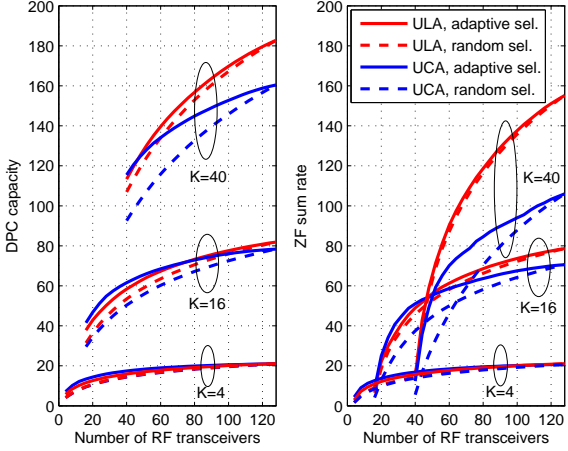} %- for double-column
  \caption{Performance of adaptive antenna selection using the convex-optimization scheme,
	as compared to performance of random selection,
  when four users are distributed at MS~5-8, and sixteen and forty users are distributed at MS~1-8.
	``ULA'' and ``UCA'' stand for uniform linear array and uniform cylindrical array, respectively.
  %For sixteen users ($K\!=\!16$) and 60 active transceivers, 
	%we gain 6\% and 14-17\% when performing adaptive antenna selection with the linear and cylindrical arrays, 
	%respectively, compared to random selection.
  %To reach 90\% of the full MIMO performance, 
	%more than 80 and 70 RF transceivers are required for the two arrays, respectively. 
  %For forty users ($K\!=\!40$), we gain 4\% and 16-50\% with the two arrays,
	%respectively,
	%and 90 and 80 transceivers are needed to reach 90\% of the full MIMO performance.
  %In the worst case, i.e., the linear array serving forty users, we can switch off up to 40 transceivers.
  }
  \label{fig:dpc_zf_rate_more_users}
\end{figure}

The antenna selection performance is shown in Fig.~\ref{fig:dpc_zf_rate_more_users}.
The case of four users from Fig.~\ref{fig:ant_sel_gain_nlos_sp_fix_tx_pow} is also included for comparison.
When the number of users increases, the DPC capacity increases.
With ZF precoding, however, the sum-rate for more users can be lower,
i.e., when the number of active transceivers is close to the number of users.
For example, with 16 antennas, the sum-rate with 16 users is lower than with four users
and with 40 antennas, the sum-rate for 40 users is also lower than with  four users.
This is due to high inter-user interference when the number of active antennas is not large enough to spatially separate the users.
In this case, ZF has to waste a large amount of power on nulling the user interference.
With more transceivers being switched on, user interference reduces and ZF sum-rate increases.

We now investigate  how much we gain by adaptive antenna selection, as compared to random selection.
If we draw a vertical line at 60 RF transceivers, 
we can see that for sixteen users and the linear array, we gain 6\% with both DPC and ZF,
while for the cylindrical array, we gain 14\% and 17\% with DPC and ZF, respectively.
For forty users and the linear array, we gain 4\% with both DPC and ZF,
while for the cylindrical array, we gain 16\% and 50\% with DPC and ZF, respectively.
With more users, we do not gain much by doing adaptive antenna selection for the linear array,
however, for the cylindrical array, antenna selection helps significantly in improving the performance.
With sixteen users, to reach 90\% of full MIMO performance, the linear array needs more than 80 RF transceivers,
while the cylindrical array needs more than 70.
With forty users, 90 and 80 RF transceivers are needed for the linear and cylindrical arrays, respectively.
Therefore, in the worst case, i.e., the linear array serving forty users, 
we can still switch off up to 40 RF transceivers.

In Fig.~\ref{fig:dpc_zf_rate_more_users}, 
we also observe that with adaptive antenna selection,
the cylindrical array outperforms the linear array marginally for relatively small numbers of active RF transceivers,
although linear array has better average performance in the case of random selection.
With sixteen users, the cylindrical array achieves higher DPC capacity when less than 80 RF transceivers are switched on,
while with forty users, the cylindrical array performs better with 40-60 active transceivers.
Then, as the number of active transceivers increases, 
the linear array becomes more and more superior.
The linear array can significantly gain from its high angular resolution,
while the cylindrical array cannot.
This can be clearly seen in the case of 40 users, 
where the linear array has much higher DPC capacity and ZF sum-rate than the cylindrical array.
These observations indicate that the compact cylindrical array can perform better for relatively small number of users,
while for larger number of users, a physically large array is preferable. 
The explanation is that for relatively small numbers of well-separated users
and active antennas,
the received SNRs at the users are more important for the performance than user channel decorrelation. 
Hence, we gain by selecting the antennas with high channel gains on the cylindrical array, 
pointing in the ``right'' directions.
This effect is even more pronounced at very low SNRs.
However, for relatively large numbers of users and active antennas, or at high SNRs,
decorrelation of user channels becomes more important for the performance than the received SNRs.
In this case, we need the high angular resolution provided by the linear array to spatially separate the users.

From the above investigations for different propagation conditions and different number of users,
we see that quite many RF transceivers can be switched off to save energy consumption and simplify massive MIMO systems, 
even when serving a relatively large number of users.
Table~\ref{table:performance_gain} and Table~\ref{table:number_of_trx_cvx} summarize the performance gain of adaptive antenna selection and the required number of RF transceivers to achieve 90\% of full MIMO performance.
Next, we use these antenna selection results based on convex optimization as a benchmark,
and evaluate how well the simple power-based selection scheme performs.

\begin{table}
\footnotesize
\begin{center}
		\caption{The performance gain by performing antenna selection, as compared to random selection.}
		\label{table:performance_gain}
    \begin{tabular}{l | l | l | l}
    \hline
    No. of users & Scenario & \multicolumn{2}{c}{Performance gain\textsuperscript{1}} \\
    \cline{3-4}
                 &                       & Linear array & Cylindrical array \\ 
    \hline
    4   & Co-located, LOS   & 11-18\%      & $>$30\% \\ \cline{2-4}
        & Far apart, NLOS   & 10\%         & 20\% \\ \hline
    16  & Mixed\textsuperscript{2} & 6\%			 & 14-17\% \\ \hline
    40  & Mixed\textsuperscript{2} & 4\%       & 16-50\% \\
    \hline
    \end{tabular}
\end{center}
\textsuperscript{1} The performance gain in terms of DPC and ZF sum-rates are at 40 RF transceivers for 4 users, and at 60 transceivers for 16 and 40 users.\\
\textsuperscript{2} ``Mixed'' means that among the 16 or 40 users some are co-located at the same site,
while users at different sites have large spacing. Half of the users are in LOS and half are in NLOS.
\end{table}

\begin{table}
\footnotesize
\begin{center}
		\caption{The required number of RF transceivers to achieve 90\% of full MIMO performance, with the convex-optimization selection scheme.}
		\label{table:number_of_trx_cvx}
    \begin{tabular}{l | l | l | l}
    \hline
    No. of users & Scenario & \multicolumn{2}{c}{No. of RF transceivers} \\
    \cline{3-4}
                 &                       & Linear array & Cylindrical array \\ 
    \hline
    4   & Co-located, LOS   & 70           & 50-60 \\ \cline{2-4}
        & Far apart, NLOS   & 80      & 60 \\ \hline
    16  & Mixed\textsuperscript{1} & 80		 & 70 \\ \hline
    40  & Mixed\textsuperscript{1} & 90    & 80 \\
    \hline
    \end{tabular}
\end{center}
\textsuperscript{1} ``Mixed'' means that among the 16 or 40 users some are co-located at the same site,
while users at different sites have large spacing. Half of the users are in LOS and half are in NLOS.
\end{table}

\subsection{Performance of Power-Based Antenna Selection}

For spatial multiplexing in multi-user MIMO systems, 
the goal is to have separated data streams to different users, 
it is thus not optimal to use the signals from two highly-correlated antennas, even if both have high SNRs.
%as discussed in Sec.~\ref{sec:intro}.
To obtain optimal antenna combinations,
there is a trade off between antenna channel correlation and SNR,
as what is done in the near-optimal convex-optimization selection scheme.
In the power-based selection scheme, antenna correlation is not considered.
The performance of this simple scheme thus depends on 
whether the antenna channels are highly correlated or not.
In the scenario where users are co-located with LOS,
the correlation between antenna channels are higher, 
as compared to the NLOS scenario with well-separated users.
We therefore expect the power-based scheme to work better in the latter case.

\begin{figure}
  \centering
  \includegraphics[width=0.48\textwidth]{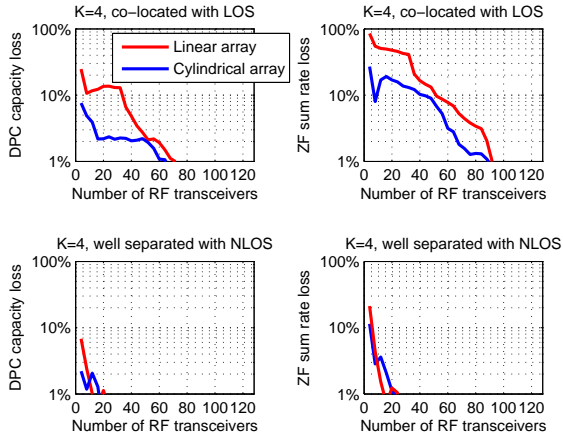} %- for double-column
  \caption{Performance loss of the power-based antenna selection scheme, relative to the near-optimal convex-optimization scheme, 
  in the scenarios where four users are closely located at MS~2 with LOS conditions, and four users are well separated at MS~5-8 with NLOS conditions.
  %In the former scenario, the loss in DPC capacity and ZF sum-rate is below 1\% when we have 70 and 90 active RF transceivers, respectively.
  %In the latter scenario, already at 20 active RF transceivers, the performance loss is below 1\%.
	}
  \label{fig:ant_sel_pow_loss_los_b_fix_tx_pow}
\end{figure}

In Fig.~\ref{fig:ant_sel_pow_loss_los_b_fix_tx_pow},
we show the performance loss by using the power-based selection scheme, 
relative to the performance of the convex-optimization scheme, in the two scenarios with four users, respectively.
In the figure the vertical axis is the performance loss in DPC capacity or ZF sum-rate,
and 100\% loss means that the power-based scheme gives zero capacity/sum-rate,
while a small loss, e.g., below 1\%, indicates that the power-based scheme performs very close to the convex-optimization scheme.
For co-located users with LOS,
the performance losses in both DPC capacity and ZF sum-rate are quite high,
when the number of RF transceivers is relatively small.
However, the performance losses decrease as more RF transceivers are switched on.
At around 70 RF transceivers, for both the linear and cylindrical arrays,
the loss in DPC capacity goes below 1\%.
More RF transceivers, i.e., 90 have to be switched on to reduce the ZF sum-rate loss below 1\%,
for both arrays.
For well-separated users with NLOS, as expected, 
the performance loss by using power-based scheme is much smaller, compared with the previous scenario.
Already at 20 RF transceivers,
the loss is below 1\% in both DPC capacity and ZF sum-rate.

\begin{figure}
  \centering
  \includegraphics[width=0.48\textwidth]{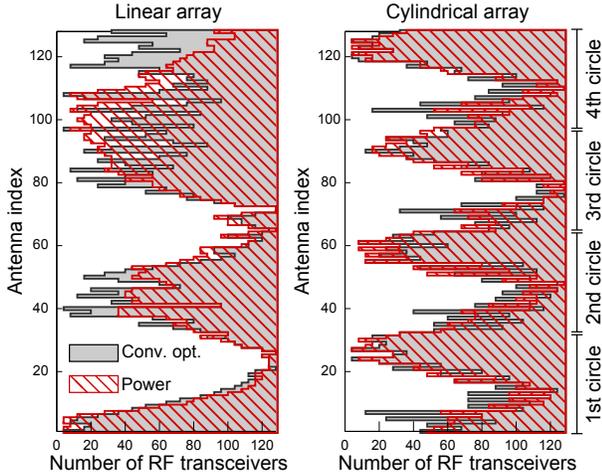} %- for double-column
  \caption{Comparison of the selected antenna indices using the convex-optimization selection scheme and the power-based selection scheme,
  as the number of active RF transceivers grows from 4 to 128.
  Four users are closely located at MS~2 and all have LOS conditions.}
  \label{fig:ant_sel_idx_subfig1}
\end{figure}
%
%\begin{figure}
%  \centering
%  \includegraphics[width=0.48\textwidth]{ant_sel_pow_loss_nlos_sp_fix_tx_pow_minus5dB}
%  \caption{Performance loss of the power-based antenna selection scheme, relative to the near-optimal convex optimization scheme, 
%  in the scenario where four users are well separated at MS~5-8 and all have NLOS conditions to the base station.}
%  \label{fig:ant_sel_pow_loss_nlos_sp_fix_tx_pow}
%\end{figure}

%In the former scenario where the four users are closely located with LOS, 
%although the power-based selection cannot perform as optimal as the convex optimation scheme due to relatively high antenna correlation at the base station array,
%we can switch on more transceivers to reduce the loss.
%With the simple selection scheme and practical ZF precoding, 90 active transceivers achieve 90\% of full MIMO performance,
%where about 40 transceivers can be switched off.
%
To better understand how the power-based scheme compares to the convex-optimization scheme,
we compare the selected antenna indices when using the two schemes.
Note that the selected antenna indices are presented for a single coherence interval,
and the indices may change over time due to fading.
The comparison is shown in Fig.~\ref{fig:ant_sel_idx_subfig1} for both arrays, 
in the more ``difficult'' scenario where the four users are co-located with LOS.
We can see why the power-based scheme performs worse for smaller number of active transceivers.
With less than 60 RF transceivers, 
the antenna indices selected by the two schemes are quite distinct, for both arrays.
With more RF transceivers, 
the difference becomes smaller and smaller, and eventually vanishes when all antennas are used.

We first focus on the differences on the linear array.
Note that the shape of the selected antenna indices by the power-based scheme
is similar to the power variation over the array shown in Fig.~\ref{fig:APS_both}(c)
where one user is at MS~2. 
The antennas at indices about 1-10, 40-50 and 80-110 are favored by the power-based scheme, 
due to the power contribution from the LOS component and the significant scatterers, as shown in Fig.~\ref{fig:APS_both}(a).
With about 20 to 60 active transceivers, 
the power-based scheme selects the neighboring antennas at index about 80-110 that have relatively high channel gain,
while the convex-optimization scheme avoids selecting those neighboring ones at the same time due to their highly correlated channels.
The convex-optimization scheme trades off between antenna correlation and channel gain,
and selects the antennas at index around 40 and 120 instead.
Considering the influence on capacity from the two selections,
capacity increases logarithmically with SNR and
linearly with the number of orthogonal spatial dimensions.
At a relatively high SNR, it is preferable to select uncorrelated channels, 
which contribute to all spatial dimensions.
This effect can be observed in the performance loss of the power-based scheme
with the linear array in Fig.~\ref{fig:ant_sel_pow_loss_los_b_fix_tx_pow}.
Above 30 RF transceivers the performance loss starts to decrease rapidly.
This is because the power-based scheme starts to select antennas with index around 40 that have lower correlation with those at index 80-110,
and the performance of the power-based scheme is boosted.
We can also explain why ZF precoding needs more active transceivers to have a performance loss below 1\%.
The ZF precoding is more sensitive to user interference than the DPC,
and the power-based scheme first selects antennas with high channel gain but high correlation,
which is not preferable for reducing the user interference.

The antenna indices on the cylindrical array are ordered 
from the bottom circle (the 1st circle) to the top circle (the 4th circle) on the array.
In each circle, the first antenna is pointing north (up in Fig.~\ref{fig:MAP_small}),
and the antenna indices are ordered counter-clockwise.
In Fig.~\ref{fig:ant_sel_idx_subfig1},
the selected antenna indices clearly show the four-circle structure,
as well as the dual polarization since every other antennas are selected first.
On each circle, the antennas pointing in the direction of MS~2, where the users are located,
are selected first by the power-based scheme.
This is clearly seen when there are about 20-50 RF chains.
%With about 20 to 50 RF transceivers, 
%the antennas pointing in the same direction, i.e., the user direction of MS~2, on the four circles are selected first by the power-based scheme.
However, these antennas are closely-spaced in elevation and thus experience higher channel correlation.
%the closely-spaced antennas in elevation have higher correlation,
This effect can be observed in Fig.~\ref{fig:ant_sel_pow_loss_los_b_fix_tx_pow} with the cylindrical array,
where the performance loss decreases slowly between 20 and 50 RF transceivers.
Above 50 RF transceivers, the antennas pointing in other directions and with lower correlations start to be selected, 
therefore, the performance loss drops quickly.

\begin{figure}
  \centering
  \includegraphics[width=0.48\textwidth]{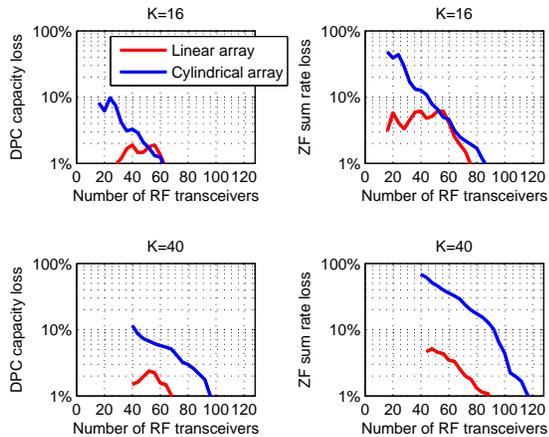} %- for double-column
  \caption{Performance loss of the power-based antenna selection scheme, relative to the near-optimal convex-optimization scheme, 
  in the cases of sixteen and forty users.
  %For sixteen users, 60 and 80 active RF transceivers are required to have the performance loss below 1\%, with DPC and ZF, respectively.
  %For forty users, 70 and 90 are needed with the linear array, 100 and 120 are needed with the cylindrical array, with DPC and ZF, respectively.
	}
  \label{fig:ant_sel_pow_loss_more_users_fix_tx_pow}
\end{figure}

We move on to the cases of more users,
shown in Fig.~\ref{fig:ant_sel_pow_loss_more_users_fix_tx_pow}.
For sixteen users and both arrays,
we need to switch on 60 transceivers to make the performance loss below 1\%, with DPC.
With ZF, more than 80 transceivers are needed.
The fluctuations in the performance loss for the linear array can be explained
by the fact that above 60 RF chains the power-based scheme starts to select antennas with lower correlations,
thus the performance loss drops quickly then.
Moving up to forty users, an even larger number of transceivers has to be switched on.
With the linear array, about 70 and 90 transceivers can make the performance loss below 1\% in DPC capacity and ZF sum-rate, respectively. 
With the cylindrical array, up to 100 transceivers are needed with DPC, while 120 are needed with ZF precoding.
This indicates that we need almost all the transceivers with the cylindrical array, for forty users, when using ZF.
If we accept a bit lower sum-rate by the power-based selection scheme, 
e.g., allowing 5\% loss,
we can reduce the required number of transceivers to 100.

%
%\begin{figure*}[h!t]
%	\centering
%	%\hspace{0.1cm}
%	\subfigure[]{
%		\includegraphics[scale=0.38]{ant_sel_idx_4_users_los_minus5dB}
%	  \label{fig:ant_sel_idx_subfig1}
%	 } 
%%	\subfigure[]{
%%		\includegraphics[scale=0.38]{ant_sel_idx_4_users_nlos_minus5dB}
%%	  \label{fig:ant_sel_idx_subfig2}http://open.spotify.com/album/6izAbH73yKhuYP0QPVIyQ1
%%	 }
%	\subfigure[]{
%		\includegraphics[scale=0.38]{ant_sel_idx_40_users_los_nlos_minus5dB}
%	  \label{fig:ant_sel_idx_subfig3}
%	 }
%	\caption{Comparison of selected antenna indices when using the convex optimization scheme and the power-based scheme.}
%	\label{fig:ant_sel_idx}
%\end{figure*}

\begin{figure}
  \centering
  \includegraphics[width=0.48\textwidth]{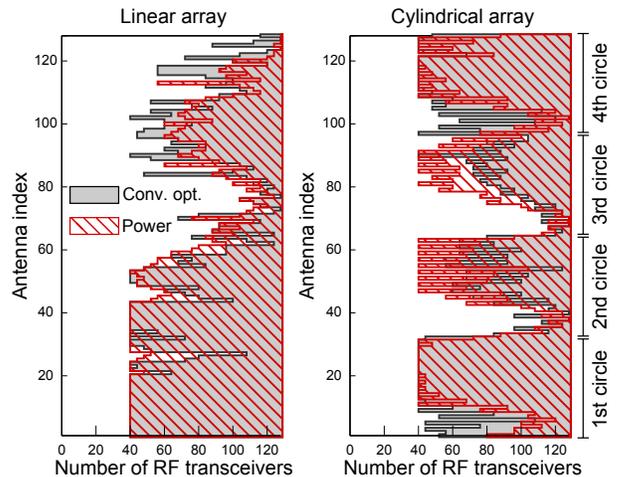} %- for double-column
  \caption{Comparison of the selected antenna indices using the convex-optimization selection scheme and the power-based selection scheme,
  as the number of active RF transceivers grows from 40 to 128.
  Forty users are distributed at MS~1-8.}
  \label{fig:ant_sel_idx_subfig3}
\end{figure}

In Fig.~\ref{fig:ant_sel_idx_subfig3}, we show the difference in antenna indices selected by the two schemes, when there are 40 users.
Again, the selected antenna indices are presented for a single coherence interval.
For the linear array, the power-based scheme selects the antennas at index 1-60 due to high channel gain,
while the convex-optimization scheme selects some antennas at index around 100 instead so that high antenna correlation can be avoided.
When more than 70 transceivers are switched on, the power-based scheme starts to select antennas at index around 100,
the performance gap between the two schemes drops below 1\%, as shown in Fig.~\ref{fig:ant_sel_pow_loss_more_users_fix_tx_pow}.

For the cylindrical array, the difference in the selected antenna indices is significant.
The power-based scheme first selects antennas facing the user directions on each circle,
however, those antennas have high correlation due to their small separation in both azimuth and elevation.
The convex-optimization scheme tries to split them and selects antennas on the 1st and 4th circles instead,
although some of the antennas are not pointing in the user directions.
Due to the large number of users in this case, 
the power-based scheme needs more antennas to spatially separate the users and achieve performance close to the convex-optimization scheme,
especially with ZF.
%This is a main reason for the power-based scheme not performing very well for the cylindrical array in this case. 
%\todo[inline]{I think the text above is a bit confusing too. A re-write may be in order.\\Xiang: have re-written.}

From all above observations,
the power-based selection scheme gives very competitive results.
Only in those ``difficult'' scenarios, such as closely spaced users with LOS, 
and a relatively high number of users served by the cylindrical array,
we need more antennas and transceivers to achieve performance close to the convex-optimization scheme.
To summarize, Table~\ref{table:number_of_trx} lists the number of required RF transceivers that achieves 90\% of full MIMO performance,
when using the simple power-based selection scheme and practical ZF precoding.
From there, we see that generally a large number of RF transceivers can be switched off.
This opens up an opportunity to apply this simple antenna selection scheme in massive MIMO.

\begin{table}
\footnotesize
\begin{center}
		\caption{The required number of RF transceivers, with the simple power-based selection scheme and practical ZF precoding.}
		\label{table:number_of_trx}
    \begin{tabular}{l | l | l | l}
    \hline
    No. of users & Scenario & \multicolumn{2}{c}{No. of RF transceivers} \\
    \cline{3-4}
                 &                       & Linear array & Cylindrical array \\ 
    \hline
    4   & Co-located, LOS   & 90           & 90 \\ \cline{2-4}
        & Far apart, NLOS   & 80           & 60 \\ \hline
    16  & Mixed\textsuperscript{1} & 80					 & 80 \\ \hline
    40  & Mixed\textsuperscript{1} & 90           & 120 \\
    \hline
    \end{tabular}
\end{center}
\textsuperscript{1} ``Mixed'' means that among the 16 or 40 users some are co-located at the same site,
while users at different sites have large spacing. Half of the users are in LOS and half are in NLOS.
\end{table}

The impact of SNR on antenna selection should be mentioned here.
At very low SNRs, it is more preferable to select antennas with high channel gains 
so that array gain can be achieved to boost the capacity,
while at high SNRs, better user separation is more important for spatial multiplexing.
Therefore, more RF transceivers should be switched on at very low SNRs than at higher SNRs. 
%In our investigations above, we set $\rho\!=\!-5$ dB, which causes a relatively high receive SNR at the users,  
%especially when lots of transceivers are switched on.
%With high SNR, more RF transceivers can be reduced due to the reason that MIMO capacity increases with SNR in logarithm,
%as compared to very low SNR.
%At very low SNR, array gain is important to boost the capacity,
%thus more transceivers should be switched on.
%\todo[inline]{This explanation is difficult to follow. Maybe the argumentation can be reversed to make it more clear. Starting with the fact that at low SNRs it is more important to include antennas to achieve antenna gain, while at high SNRs it is only needed if better user separation can be achieved ... or something similar.}
%High SNR: high spectral efficiency, more RF tranceivers can be redueced due to the fact that some antennas do not contribute much to spectral efficiency.
%But high SNR will cause hardware problem at receiver, e.g., power amplifier, AGC...
%Large constellations can be used for transmission.
%Low SNR: lower spectral efficiency, less RF tranceivers can be reduced due to that array gain is important here to boost spectral efficiency.
%This is the advantage of massive MIMO that gains from array gain, and is more favorable.
%But when hardware power consumption needs to be considered, using more transceivers will cause the decrease of EE.
%It would be intesting to measure the hardware power consumption and study the optimal number of transceivers.
%With lower SNR, power-based selection will be more closer to the optimal.
However, considering hardware power consumption, 
more active transceivers may decrease the overall energy efficiency of massive MIMO. 
It would be interesting to measure hardware power consumption at the base station, 
and investigate optimal amount of transmit power and optimal number of active transceivers that 
maximizes energy efficiency; however this has to be left for future work.

\section{Summary and Conclusions}\label{sec:conclusions}

Unlike the situation in i.i.d.\ Rayleigh channels, where all antennas contribute equally,
in real propagation channels, large-scale fading over the arrays or differences in antenna patterns, 
makes some antennas contribute more than others.
Using channels measured at 2.6~GHz with a linear array with omni-directional elements and a cylindrical array with patch elements,
we have illustrated that 
a significant performance gain can be achieved by performing adaptive antenna selection,
as compared to random selection. 
A substantial number of RF transceiver chains can be turned off without a significant performance loss.
Antenna selection based on a convex-optimization scheme gives near-optimal performance, however,
a very simple selection scheme that is only based on received signal power measurements at each antenna also gives very competitive results.

The overall conclusion from our work is that antenna selection may be effectively used to reduce the implementation complexity, 
cost and hardware energy consumption of massive MIMO systems.
The difference in characteristics between theoretical i.i.d.\ Rayleigh and real propagation channels also underlines
the importance of developing new channel models for massive MIMO.

\section*{Acknowledgement}
The authors would like to acknowledge the support from ELLIIT - an Excellence Center at Link{\"o}ping-Lund in Information Technology,
the Swedish Research Council (VR), and the Swedish Foundation for Strategic Research (SSF).
The research leading to these results has received funding from the European Union Seventh Framework Programme (FP7/2007-2013)
under grant agreement n$^{\circ}$ 619086 (MAMMOET).

\bibliographystyle{IEEEtran}
\bibliography{bib/IEEEabrv,bib/antenna_selection_refs}

\end{document}